\documentclass[lonbibliography, reprint, aps, prd, superscriptaddress, nofootinbib, floatfix]{revtex4-2} 
\usepackage{amsmath, amssymb, mathtools, graphicx, microtype}\allowdisplaybreaks 
\usepackage[noautoscale]{youngtab}\Yboxdim 6.5pt\Ylinethick 0.5pt
\usepackage[bb=dsfontserif]{mathalpha}
\usepackage[dvipsnames]{xcolor} 
\usepackage{orcidlink}  
\usepackage[framemethod=TikZ, backgroundcolor=Aquamarine!15]{mdframed}
\mdfsetup{skipabove=0pt, skipbelow=0pt, linewidth=0pt}  
\usepackage{hyperref}
\graphicspath{{Figures/}}
 
\newcommand{\T}{\mathbb{T}} 
\newcommand{\Z}{\mathbb{Z}}  

\newcommand{\U}{\textrm{U}}
\newcommand{\SU}{\textrm{SU}}
\newcommand{\USp}{\textrm{USp}} 
 
\begin{document}

\title{Fermion Masses and Mixings in String Theory with Dirac Neutrinos}

\author{Mudassar Sabir\,\orcidlink{0000-0002-8551-2608}}
\email{mudassar.sabir@uestc.edu.cn} 
\affiliation{School of Physics, University of Electronic Science and Technology of China, Sichuan 611731, Chengdu, China}

\author{Adeel Mansha\,\orcidlink{0000-0002-1183-0355}}
\email{adeelmansha@alumni.itp.ac.cn}
\affiliation{College of Physics and Optoelectronic Engineering, Shenzhen University, Shenzhen 518060, China}

\author{Tianjun Li\,\orcidlink{0000-0003-1583-5935}}
\email{tli@itp.ac.cn}
\affiliation{School of Physics, Henan Normal University, Xinxiang 453007, China}
\affiliation{School of Physical Sciences, University of Chinese Academy of Sciences, Beijing 100190, China}
\affiliation{CAS Key Laboratory of Theoretical Physics, Institute of Theoretical Physics, Chinese Academy of Sciences, Beijing 100190, China}

\author{Zhi-Wei Wang\,\orcidlink{000-0002-5602-6897}}%\thanks{Corresponding author}
\email{zhiwei.wang@uestc.edu.cn}
\affiliation{School of Physics, University of Electronic Science and Technology of China, Sichuan 611731, Chengdu, China}

\begin{abstract}
Analyzing the supersymmetric Pati-Salam landscape on a $\mathbb{T}^6/(\mathbb{Z}_2\times \mathbb{Z}_2)$ orientifold in IIA string theory, we have found only two models that accurately account for all standard model fermion masses and mixings. The models are dual to each other under the exchange of two SU(2) sectors and feature 12 adjoint scalars, the maximum number allowed in the landscape, whose linear combination yields the two light Higgs eigenstates. Dirac neutrino-masses in normal ordering $(50.6,~10.6,~6.2)\pm 0.1$~meV satisfying both the experimental as well as swampland constraints.   
\end{abstract}

\maketitle  

\textbf{Introduction} -- Standard Model (SM) fermions appear in chiral representations of the gauge group $\SU(3)_C \times \SU(2)_L \times \U(1)_Y$. Intersecting D6-branes in type IIA string theory provide a natural mechanism to realize chiral fermions at D-brane intersections \cite{Aldazabal:2000cn}. Family replication results from multiple intersections of D6-branes that fill four-dimensional spacetime and extend into three compact dimensions. The volumes of the cycles wrapped by D-branes determine the four-dimensional gauge couplings, while the total internal volume yields the gravitational coupling. Yukawa couplings arise from open world-sheet instantons, specifically the triangular worldsheets stretched between intersections where fields involved in the cubic coupling reside. These instanton effects are suppressed by $\exp(-A_{ijk}T)$, where $A_{ijk}$ is the area of the triangle bounded by intersections $\{i, j, k\}$ and $T$ is the string tension \cite{Cremades:2003qj}. This exponential suppression explains the fermion mass hierarchies and mixings.  

Intersecting D-branes model building with three families and realistic Yukawa textures naturally favors direct products of unitary gauge groups over the simple unitary groups. And the K-theory conditions \cite{Witten:1998cd, Uranga:2000xp}, being mod~4, are more easily satisfied for $\U(2N)$ with $N \in \Z$. Consequently, the left-right symmetric Pati-Salam group, $\SU(4)_C\times \SU(2)_L\times \SU(2)_R$, emerges as the most promising choice for realistic models. The rules to construct supersymmetric Pati-Salam models on a $\T^6/(\Z_2\times \Z_2)$ orientifold from intersecting D6-branes with the requirement of $\mathcal{N}=1$ supersymmetry, tadpole cancellation and the K-theory constraints were outlined in \cite{Cvetic:2004ui, Blumenhagen:2006ci, Blumenhagen:2005mu}. Similar construction is employed in recent works \cite{Li:2019nvi, Li:2021pxo, Mansha:2022pnd, Sabir:2022hko, Mansha:2023kwq, Mansha:2024yqz}. In ref.~\cite{He:2021gug} the complete landscape of consistent three-family supersymmetric Pati-Salam models from intersecting D6-branes on a $\T^6/(\Z_2 \times \Z_2)$ orientifold was fully mapped, comprising of 202,752 models with 33 distinct gauge-coupling relations. The viable models with realistic Yukawas split into classes of either 6, 9 or 12 adjoint scalars from $\mathcal{N}=2$ sector, whose linear combination yields the two light Higgs mass eigenstates \footnote{There also exists a model having 3 adjoint scalars from the bulk, however, tiny Yukawa couplings from the bulk Higgs are argued to be related to the infinite distance limit~\cite{Lee:2019wij} in the moduli space where a light of tower states, dubbed gonions~\cite{Aldazabal:2000cn}, appears signalling the decompactification of one or two compact dimensions \cite{Casas:2024ttx, Casas:2024clw}. Instead, the Yukawas originating from the $\mathcal{N}=2$ sector are insensitive to the bulk moduli and the issue of decompactification does not arise.}. The results for the Yukawa couplings and the analysis of soft terms from supersymmetry breaking for all viable models in the landscape are presented in \cite{Sabir:2024cgt, Sabir:2024jsx}. 

\textbf{Neutrino-sector} -- Experimentally, two of the mass eigenstates $m_1,~m_2$ are found to be close to each other while the third eigenvalue $m_3$ is separated from the former pair where $m_2 > m_1$ by definition. Normal ordering (NO) refers to $m_3\gg m_2>m_1$ while inverted ordering (IO) refers to $(m_2>m_1\gg m_3)$ with constraints \href{http://www.nu-fit.org/?q=node/294}{NuFIT 6.0 (2024)} \cite{Esteban:2024eli},
\begin{mdframed}\setlength\abovedisplayskip{0pt} 
\begin{equation}
\begin{split}\label{eq:constraints1}
\Delta m_{21}^2 &= 74.9 \pm 1.9 ~\mathrm{meV}^2,\\
\Delta m_{31}^2 &= +2513 \pm 20 ~\mathrm{meV}^2 ~ \mathrm{(NO)}, \\
\Delta m_{32}^2 &= -2484 \pm 20 ~\mathrm{meV}^2 ~ \mathrm{(IO)}, \\
\sum_{i=1}^3 m_i &> 58~(98)~\mathrm{meV} ~\mathrm{NO~(IO)}, \\
\sum_{i=1}^3 m_i &< 72~\mathrm{meV}, 
\end{split}
\end{equation}
\end{mdframed} 
where the constraint on the sum of neutrino masses is the strongest till to date from 2024 DESI BAO+CMB data \cite{DESI:2024mwx} at 95\% C.L.

Recent insights from the swampland program, particularly from the non-SUSY AdS instability conjecture \cite{Ooguri:2016pdq} and the light fermion conjecture \cite{Gonzalo:2021fma} suggests that without additional chiral fermions with tiny masses, neutrinos must be of Dirac-type together with a bound on the lightest neutrino mass given by the cosmological constant scale as, $m_{\nu}^{\rm lightest} \lesssim \Lambda^{1/4}$. The 3D Casimir energy of the SM compactified on a circle receives a positive contribution from the lightest neutrino, which is necessary to avoid unstable non-supersymmetric AdS vacua. This constraint is only satisfied for Dirac neutrinos, which carry 4 degrees of freedom, unlike Majorana neutrinos, which only have 2 and cannot compensate for the 4 bosonic degrees of freedom from the photon and the graviton. This also avoids the inevitable lepton-number violations in the Majorana case. In refs.~\cite{Arkani-Hamed:2007ryu, Arnold:2010qz, Ibanez:2017kvh, Hamada:2017yji, Gonzalo:2018tpb, Gonzalo:2021zsp, Castellano:2023qhp, Anchordoqui:2023wkm} the 3D Casimir energies corresponding to the compactification of the standard model on a circle were computed resulting in the following bounds: 
\begin{mdframed}\setlength\abovedisplayskip{0pt} 
\begin{align}
m_{\nu}^{\rm lightest} & < 7.7~(2.6)\pm 0.5~\mathrm{meV}~ \mathrm{NO~(IO)} ~\mathrm{(Dirac)} \nonumber\\
\sum_{i=1}^3 m_i &= 65~(105) \pm 5 ~\mathrm{meV}~ \mathrm{NO~(IO)} ~\mathrm{(Dirac)}  \label{eq:constraints2}
\end{align} 
\end{mdframed}  
where the last constraint on the sum of neutrino-masses comes from applying the multiple point criticality principle \footnote{In analogy with the first-order transition between ice and water the slush exists at $0^\circ$C. Conversely, if the temperature happens to be close to zero, it is because of the existence of such as a slush \cite{Froggatt:1995rt}}. Since the AdS and the dS vacua are separated by infinite distance in the moduli-space \cite{Lust:2019zwm}, any transition between them is of first-order. The multiple point criticality principle thus requires the 3D dS vacuum to be close to the flat vacuum \cite{Hamada:2017yji}. 

Henceforth, it is crucial in string theory to generate tiny Dirac Yukawa couplings while keeping the other Yukawa couplings and SM gauge couplings unsuppressed. Previous efforts to generate tiny neutrino-masses have focused on Euclidean D2-brane instantons within local models without realistic Yukawa textures \cite{Blumenhagen:2009qh, Cvetic:2008hi, Ibanez:2008my}, see Ref.~\cite{Casas:2024clw} for a recent survey on this issue.

\begin{table}[t]\footnotesize
	\caption{D6-brane configurations and intersection numbers of Model~\hyperref[model22]{22}, and its MSSM gauge coupling relation is $g^2_a=\frac{5}{6}g^2_b=\frac{11}{6}g^2_c=\frac{11}{8}(\frac{5}{3}g^2_Y)=\frac{8}{7 \sqrt{3}} \sqrt[4]{2}\, 5^{3/4}\, \pi \,e^{\phi_4}$. 
%Model 22-dual is obtained by exchanging $b\leftrightarrow c$ with the gauge coupling relation $g^2_a=\frac{11}{6}g^2_b=\frac{5}{6}g^2_c=\frac{25}{28}(\frac{5}{3}g^2_Y)=\frac{8}{7 \sqrt{3}}\, \sqrt[4]{2}\, 5^{3/4}\, \pi \,e^{\phi_4}$.
}
$\begin{array}{|c|c|c|c|c|c|c|c|c|c|}
\hline\multicolumn{2}{|c|}{\text{Model~\hyperref[model22]{22}}} & \multicolumn{8}{c|}{ {\SU(4)_C\times \SU(2)_L \times \SU(2)_R\times \USp(2)} }\\
\hline \hline\rm{stack} & N & (n^1,l^1)\times(n^2,l^2)\times(n^3,l^3) & n_{\yng(2)}& n_{\yng(1,1)_{}} & b & b' & c & c' & 3\\
\hline
a & 8 & (1, -1)\times (1, 0)\times (1, 1) & 0 & 0  & 3 & 0 & -3 & 0 & 0\\
b & 4 &  (-2, 5)\times (0, 1)\times (-1, 1) & 3 & -3  & \text{-} & \text{-} & 0 & -8 & 2\\
c & 4 &  (2, 1)\times (1, 1)\times (1, -1) & 2 & 6  & \text{-} & \text{-} & \text{-} & \text{-} & 2\\   
3 & 2  &  (0, -1)\times (1, 0)\times (0, 2) & \text{} & \text{} & \text{} & \text{} & \text{} & \text{} & \\ 
\hline
% &   &    \multicolumn{8}{c|}{ \beta^g_3=-2, ~ \chi_1=\sqrt{\frac{2}{5}} , ~ \chi_2=\frac{\sqrt{\frac{5}{2}}}{6} , ~  \chi_3=2 \sqrt{\frac{2}{5}} }\\
%\hline  
\end{array}$ \label{model22} 
\end{table}

\textbf{The Pati-Salam model} -- In this letter, we present the only two models in the supersymmetric Pati-Salam landscape from intersecting D6-branes on a $\T^6/(\Z_2\times \Z_2)$ orientifold that accurately accommodate all standard model fermion masses and mixings, while also providing a unique prediction for the Dirac-neutrino masses. This necessitates the inclusion of at least twelve adjoint scalars from the $\mathcal{N}=2$ sector, which is the maximum available in the landscape \cite{Sabir:2024cgt}. The two light Higgs eigenstates arise from the linear combination of the vacuum expectation values (VEVs) $v^i_{u,d} = \langle H^i_{u,d}\rangle$ of the twelve adjoint Higgs present in the model. 

Majorana-neutrino masses can always be added via the type-I seesaw mechanism taking Dirac-neutrino mass matrix as an input \cite{Mayes:2019isy}, whereby the right-handed neutrino masses can be generated via the stringy instanton effects~\cite{Blumenhagen:2006xt, Ibanez:2006da, Cvetic:2007ku}. To evade the AdS vacua in the case of Majorana neutrinos, the model has 9 SM singlet chiral supermultiplets from the $\SU(2)_L$ and $\SU(2)_R$ antisymmetric representations which can play the roles similar to the sterile neutrinos~\cite{Ibanez:2017kvh}. Here, we only focus on the minimal simplest case with tiny Dirac-neutrinos. 

Table \ref{model22} displays the intersection numbers among the three D6-brane sectors ($a,b,c$) and an O6-plane sector (3) in the model. The dual model is constructed by exchanging the two SU(2) stacks $b$ and $c$. Pati-Salam gauge symmetry ${\rm SU}(4)_C\times {\rm SU}(2)_L\times {\rm SU}(2)_R$ is higgsed down to the SM gauge group ${\rm SU}(3)_C\times {\rm U}(2)_L\times {\rm U}(1)_{I3R}\times {\rm U}(1)_{B-L}$ by assigning vacuum expectation values (VEVs) to the adjoint scalars which arise as open-string moduli associated to the stacks $a$ and $c$. Moreover, the ${\rm U}(1)_{I_{3R}}\times {\rm U}(1)_{B-L}$ gauge symmetry may be broken to ${\rm U}(1)_Y$ by giving VEVs to the vector-like particles with the quantum numbers $({ 1, 1, 1/2, -1})$ and $({ 1, 1, -1/2, 1})$ under the ${\rm SU}(3)_C\times {\rm SU}(2)_L\times {\rm U}(1)_{I_{3R}} \times {\rm U}(1)_{B-L} $ gauge symmetry \cite{Cvetic:2004ui, Cvetic:2004nk, Chen:2006gd}. This brane-splitting results in SM quarks and leptons as,
\begin{eqnarray}
F_L(Q_L, L_L)  &\rightarrow &  Q_L + L , \nonumber \\
F_R(Q_R, L_R)  &\rightarrow &  U_R + D_R + E_R + N_R ~.
\end{eqnarray}
Similar to refs.~\cite{Cvetic:2007ku, Chen:2007zu} we can decouple the additional exotic particles.

\textbf{Yukawa Couplings} -- Yukawa couplings arise from open string world-sheet instantons that connect three D-brane intersections \cite{Cremades:2003qj}. 
Three-point couplings for the fermions can be read from the following superpotential,
\begin{align}\label{eq:WY3} 
\mathcal{W}_3 & \sim  Y^u_{ijk} Q_i  U^c_j H^u_k + Y^\nu_{ijk} L_i N^c_j H^u_k \nonumber\\
&\quad + Y^d_{ijk} Q_i D^c_j H^d_k +  Y^e_{ijk} L_i  E^{c}_j H^d_k ~.
\end{align} 
Yukawa couplings for D6-branes wrapping a compact $\T^2 \times \T^2 \times \T^2$ space are,
\begin{align}
Y_{ijk}& \sim  \prod_{r=1}^3 \vartheta \left[\begin{array}{c} \delta^{(r)}\\ \phi^{(r)} \end{array} \right] (\kappa^{(r)}),
\end{align}
with $r=1,2,3$ denoting the three 2-tori and the arguments of the $\vartheta$ function are,
\begin{align} 
\delta^{(r)} &= \frac{i^{(r)}}{I_{ab}^{(r)}} + \frac{j^{(r)}}{I_{ca}^{(r)}} + \frac{k^{(r)}}{I_{bc}^{(r)}} + \epsilon^{(r)} + \frac{s^{(r)}}{d^{(r)}}, \nonumber\\
\phi^{(r)}   &= 0, \quad \kappa^{(r)} = \frac{J^{(r)}}{\alpha'} \frac{|I_{ab}^{(r)}I_{bc}^{(r)} I_{ca}^{(r)}|}{(d^{(r)})^2}, \label{eqn:Yinput}
\end{align}
where $d^{(r)}=g.c.d.(I_{ab}^{(r)},I_{bc}^{(r)},I_{ca}^{(r)})$, $\epsilon^{(r)}$ is a total shift that can be absorbed due to reparameterization, $s^{(r)}\equiv s^{(r)}(i,j,k) \in \Z$ is a linear function on the indices $i,j,k$ and $J$ is the \emph{complex} K\"ahler structure of the compact space $\T^2$ \cite{Cremades:2003qj}. 
%We will not consider any fluxes $\phi^{(r)}$ by setting all Wilson lines to zero i.e. $\theta_{a,b,c}^{(r)}=0$.

The diagonal mass-matrices for up-quarks, down-quarks, charged-leptons and the neutrinos with $m_t/m_b = 41.2551$ and $m_\tau/m_b = 0.0424798$ from PDG \cite{ParticleDataGroup:2024cfk} are, 
\begin{align}
D_u &= m_t \left(
\begin{array}{ccc}
 0.0000125167 & 0. & 0. \\
 0. & 0.00737672 & 0. \\
 0. & 0. & 1. \\
\end{array}
\right),\label{eq:mass-upquarks}\\
D_d &= m_b \left(
\begin{array}{ccc}
 0.0011236 & 0. & 0. \\
 0. & 0.0223524 & 0. \\
 0. & 0. & 1. \\
\end{array}
\right) \label{eq:mass-downquarks},\\
D_e &= m_{\tau} \left(
\begin{array}{ccc}
 0.00287574 & 0. & 0. \\
 0. & 0.594612 & 0. \\
 0. & 0. & 1. \\
\end{array}
\right), \label{eq:mass-chargedleptons}\\
D_\nu &= m_{\nu} \left(\begin{array}{ccc}
m_3 & 0 & 0  \\
0 & m_2 & 0 \\
0 & 0 & m_1
\end{array} \right),  \label{eq:mass-neutrinos}
\end{align} 
where we have parameterized the neutrino-masses as $(m_3,m_2,m_1)$ upto an overall constant $m_\nu$.   
Employing the quarks-mixing matrix, $V_{\rm CKM}$, from \href{http://www.utfit.org/UTfit/ResultsSummer2023SM}{UTfit (2023)} \cite{UTfit:2022hsi} and the leptons-mixing matrix, $U_{\rm PMNS}$ from \href{http://www.nu-fit.org/?q=node/294}{NuFIT}, we express the up-quark matrix and the charged-leptons matrix in the mixed form as, \cite{Sabir:2022hko}   
\begin{widetext}
\begin{align}
M_u &= V_{\rm CKM}^{\dag} D_u V_{\rm CKM} \nonumber\\
&=m_t \left(
\begin{array}{ccc}
 0.000458944        & 0.0019479 e^{i 0.0716746} & 0.00863679 e^{i 0.387353} \\
 0.0019479 e^{-i0.0716746} & 0.00868911        & 0.0414746 e^{-i0.0181756} \\
 0.00863679 e^{-i0.387353} & 0.0414746 e^{i 0.0181756} & 0.99824        \\
\end{array}
\right) ,\label{eq:mixing-quarks} \\
M_e &= U_{\rm PMNS} D_e U_{\rm PMNS}^{\dag} \nonumber\\
& = m_\tau \left(
\begin{array}{ccc}
 0.287128        & 0.221395 e^{i 0.37897} & 0.249054 e^{i 2.62527} \\
 0.221395 e^{-i 0.37897} & 0.553552        & 0.315556 e^{i 0.109961} \\
 0.249054 e^{-i 2.62527} & 0.315556 e^{-i 0.109961} & 0.756807        \\
\end{array}
\right) . \label{eq:mixing-chargedleptons}
\end{align}
\end{widetext}
To explain all SM fermion masses and mixings, we only need to fit $\{M_u,D_d,D_\nu,M_e\}$ by solving for the 24 Higgs VEVs and arguments $\{\kappa^{(1)},
\epsilon_{3u}^{(1)},\epsilon_{3d}^{(1)},\epsilon_{3\nu}^{(1)},\epsilon_{3e}^{(1)}\}$ of theta function that fix all Yukawa couplings.
   
\textbf{Three-Point Mass Matrices} -- From table~\ref{model22}, the relevant intersection numbers are,
\begin{align}
I_{ab}^{(1)} &=3,     & I_{bc}^{(1)} &=-12, & I_{ca}^{(1)} &=-3, \nonumber\\
I_{bb'}^{(1)} &= -20, & I_{cc'}^{(1)}&= 4, & I_{bc'}^{(1)} &=-8 \label{eq:Intersections}
\end{align}
Three-point Yukawa couplings arise from the triplet intersections from the branes ${a, b, c}$ on the first two-torus ($r=1$) with 12 pairs of Higgs from $\mathcal{N}=2$ sector. Yukawa matrices for the Model~\hyperref[model22]{22} are of rank 3 and the three intersections required to form the disk diagrams for the Yukawa couplings all occur on the first torus. The other two-tori only contribute an overall constant that has no effect in computing the fermion mass ratios. Thus, it is sufficient for our purpose to only focus on the first torus.
The characteristics and the argument of the modular theta function as defined in \eqref{eqn:Yinput} become,
\begin{align}
\delta^{(1)} &= \frac{i^{(1)}}{3} - \frac{j^{(1)}}{3} - \frac{k^{(1)}}{12}+ \frac{s^{(1)}}{3}, \nonumber \\
\phi^{(1)} &=0 ,  \qquad \kappa^{(1)} = \frac{12 J^{\text{(1)}}}{\alpha '},  \label{eq:kappa.22}
\end{align}
where $i=\{0,\dots ,2\}$, $j=\{0,\dots ,2\}$ and $k=\{0,\dots ,11\}$ which respectively index the left-handed fermions, the right-handed fermions and the Higgs fields.  

The selection rule for the occurrence of a trilinear Yukawa coupling for a given set of indices is,
\begin{equation}\label{selection-rule.22}
i^{(1)} + j^{(1)} + k^{(1)} = 0 \mathrm{~mod~} 3.
\end{equation}
Then the rank-3 mass-matrix for the fermions can be determined by taking shift $s^{(1)}=j$ in \eqref{eq:kappa.22}, 
\begin{widetext}
\begin{align}
Z_{3} &= \left(
\begin{array}{ccc}
 T_0 v_1+T_9 v_4+T_6 v_7+T_3 v_{10} & T_{10} v_3+T_7 v_6+T_4 v_9+T_1 v_{12} & T_{11} v_2+T_8 v_5+T_5 v_8+T_2 v_{11} \\
 T_2 v_3+T_{11} v_6+T_8 v_9+T_5 v_{12} & T_3 v_2+T_0 v_5+T_9 v_8+T_6 v_{11} & T_4 v_1+T_1 v_4+T_{10} v_7+T_7 v_{10} \\
 T_7 v_2+T_4 v_5+T_1 v_8+T_{10} v_{11} & T_8 v_1+T_5 v_4+T_2 v_7+T_{11} v_{10} & T_6 v_3+T_3 v_6+T_0 v_9+T_9 v_{12} \\
\end{array}
\right),\label{eq:s-i22}
\end{align} 
\end{widetext}
where $v_i = \left\langle H_{i} \right\rangle$ and the three-point coupling functions are given in terms of Jacobi-theta function, 
\begin{align}\label{eq:3couplings22}
T_k  &\equiv  \vartheta \left[\begin{array}{c}
\epsilon^{(1)}_3+\frac{k}{12}\\  0 \end{array} \right]
(\kappa^{\text{(1)}}),  \quad   k={0,\cdots,11}.
\end{align} 
The couplings functions \eqref{eq:3couplings22} in the four mass-matrices $\{Z_{3u},Z_{3d},Z_{3\nu},Z_{3e}\}$ are all fixed by setting,
\begin{align}\label{eq:brane-positions}
\epsilon^{(1)}_{3u} = \epsilon^{(1)}_{3d} = 0, \quad \epsilon^{(1)}_{3e} = \epsilon^{(1)}_{3\nu} = \frac{1}{2}, \quad \kappa^{(1)} = 66 .
\end{align}
Since modular-theta function is quasi-doubly periodic, the choice of 0 or 1/2 for $\epsilon^{(1)}$'s is natural for stabilizing the open-string moduli, while $\kappa^{(1)}$ is fixed by the experimental constraints \eqref{eq:constraints1}.

The 12 up-type Higgs VEVs $v^u_i =v^\nu_i$ are then determined by fitting the 9 entries in $M_u$ and 3 off-diagonal zeros in $D_\nu$. Similarly, 9 out of 12 down-type Higgs VEVs $v^d_i =v^e_i$ are determined by $D_d$ and 3 can either be determined by the diagonal entries in $D_e$ (if we ignore the lepton's mixings) or the diagonal entries in $M_e$ (if we take account of leptons' mixings and the remaining off-diagonal entries can be accounted by adding the 4-point interactions) to get,   
\begin{gather}  
\begin{array}{llllll}
 v^u_1 & = & 0.000458945 e^{i 0.000033}     & v^d_1 & = & 0.0000272353  \\
 v^u_2 & = & 0.000115601 e^{i \pi}              & v^d_2 & = & 0.0000103508 e^{i \pi } \\
 v^u_3 & = & 1.61517 e^{-i 0.0274002}           & v^d_3 & = & 0.000779274  \\
 v^u_4 & = & 6.11128\times 10^{-6} e^{-i \pi}   & v^d_4 & = & 3.93336\times 10^{-6} e^{i \pi } \\
 v^u_5 & = & 0.00868911                         & v^d_5 & = & 0.000541809  \\
 v^u_6 & = & 0.0133327 e^{i 3.05317}            & v^d_6 & = & 0.0000103675 e^{i \pi } \\
 v^u_7 & = & 13.1569 e^{i 0.0181749}            & v^d_7 & = & 0.000295652  \\
 v^u_8 & = & 0.027537 e^{i \pi/2}             & v^d_8 & = & 0.0000103508 e^{i \pi } \\
 v^u_9 & = & 0.99824                            & v^d_9 & = & 0.0242394  \\
 v^u_{10} & = & 0.00636231 e^{-i \pi/2}       & v^d_{10} & = & 3.93336\times 10^{-6} e^{i \pi } \\
 v^u_{11} & = & 2.74778 e^{-i 0.386158}         & v^d_{11} & = & 0.000778019  \\
 v^u_{12} & = & 0.0132806 e^{i \pi }            & v^d_{12} & = & 0.0000103675 e^{i \pi } \\
\end{array} \label{eq:VEVs_22}
\end{gather}
\begin{align} 
\Rightarrow  \quad Z_{3u}&= M_u , \qquad\quad  Z_{3d}= D_d,  \nonumber\\ 
|Z_{3e}|&=m_\tau \left(
\begin{array}{ccc}
 0.287128 & 0.071824 & 0.000723162 \\
 0.071824 & 0.755588 & 0.000821766 \\
 0.000723162 & 0.000821766 & 0.756807 \\
\end{array}
\right),\nonumber\\
|Z_{3\nu}|&=m_\nu \left(
\begin{array}{ccc}
 13.1569 & 0 & 0 \\
 0.0003 & 2.74778 & 0 \\
 0.0065 & 0.0015 & 1.61517 \\
\end{array}
\right),\label{eq:Leptons3_22} 
\end{align}  
Hence, the quark sector is matched exactly thereby explaining all quark masses and mixings. While for the charged leptons, we present an approximate fitting for the mixed form of the leptons' matrix $|Z_{3e}|$, which will be supplemented with the 4-point contribution to account for leptons' mixings later. 

The diagonal neutrino mass-matrix $|Z_{3\nu}| \sim m_\nu (T^\nu_6v^u_7, T^\nu_6v^u_{11}, T^\nu_6v^u_3 )$ predicts neutrinos to be in normal ordering with coupling $T^\nu_6=1$ upto an overall scale $m_\nu$, to be fixed by experimental constraints. $T^\nu_6=1$ is a nice feature as it avoids extra fine-tuning, given that the neutrinos are already several orders of magnitude lighter than their quark and lepton counterparts. The experimental constraints \eqref{eq:constraints1} for the NO are satisfied by setting $m_\nu = 3.848$~meV, 
\begin{mdframed}\setlength\abovedisplayskip{0pt} 
\begin{gather} 
\Rightarrow (m_3,~ m_2,~ m_1) = (50.6,~10.6,~6.2)\pm 0.1~\mathrm{meV},\nonumber \\
 \Delta m_{21}^2 = 73.2~\mathrm{meV}^2, \quad  \Delta m_{31}^2 = +2525~\mathrm{meV}^2,\nonumber\\
 \sum_{i=1}^3 m_i = 67.4~\mathrm{meV}, ~\mathrm{(Dirac~w.~NO)}. \label{eq:neutrinos}
\end{gather}
\end{mdframed}
The prediction of Dirac-neutrino-masses is robust, as the ratios of neutrino-masses are essentially determined by the up-quarks matrix \eqref{eq:mass-upquarks} that serves as an input into the up-quarks mixing matrix \eqref{eq:mixing-quarks} given that the CKM matrix is now known with high precision. Although the uncertainties in \eqref{eq:mass-upquarks} can be significant since the unification-scale is not known precisely, however the experimental constraints \eqref{eq:constraints1} can mitigate these uncertainties. Consequently, the uncertainties in \eqref{eq:mass-upquarks} translate into the uncertainty in the K\"{a}hler modulus $\kappa^{(1)} = 66 \pm 2 $, while the overall uncertainty in neutrino-masses remains within $\pm 0.1$~meV.  
  
Comparing the results from \eqref{eq:neutrinos}, our universe avoids AdS vacua in 3D as the mass of the lightest neutrino turns out to be less than the threshold value of 7.7~meV and the sum of the masses of three Dirac-neutrino also falls within the range given by the multiple point criticality principle \eqref{eq:constraints2}.

\textbf{Leptons' Mixings from Four-Point Functions} -- The four-point couplings in Model~\hyperref[model22]{22} can come from considering interactions of ${a, b, c}$ with $b'$ or $c'$ on the first two-torus as can be seen from the intersection numbers \eqref{eq:Intersections}. There are 20 SM singlet fields $S_L^i$ and 8 Higgs-like state $H^{\prime}_{u, d}$. 
We consider four-point interactions with $b'$ with the shifts $l=\frac{k}{4} $ and $\ell=\frac{k}{3} $ taken along the index $k$ and using values from \eqref{eq:Intersections} \cite{Sabir:2024cgt},
\begin{align}\label{deltas22}
\delta &=\frac{i}{I_{ab}^{(1)}} +\frac{j}{I_{ca}^{(1)}} +\frac{k}{I_{bc}^{(1)}} +l , \nonumber\\
&= \frac{i}{3}-\frac{j}{3} ,\\
d &=\frac{\imath}{I_{b b'}^{(1)}} +\frac{\jmath}{I_{bc'}^{(1)}}+\frac{k}{I_{bc}^{(1)}} +\ell ,\nonumber\\
&= -\frac{\imath}{20}-\frac{\jmath}{8} ,
\end{align}
the matrix elements $a_{i,j,\imath}$ on the first torus from the four-point functions results in the classical 4-point contribution to the mass-matrix with VEVs $u_\imath, w_{\jmath}$ \cite{Sabir:2024cgt} and the four-point couplings given by,
\begin{align}\label{eq:4couplings22}
F_{i}  &\equiv  \vartheta \left[\begin{array}{c}
\epsilon^{(1)}_4+\frac{i}{20}\\  0 \end{array} \right]
(\kappa^{\text{(1)}}),\quad i={0,\dots,19}.
\end{align}
Since, we have already fitted the up-quarks matrix precisely, thus we set all up-type VEVs $u^u_\imath$ and $w^u_\jmath$ to be zero. Thus, we are essentially concerned with fitting charged-leptons mixing-matrix such that the corresponding corrections for the down-type quarks remain negligible. The desired solution can be readily obtained by setting $\epsilon^{(1)}_{4d}=1/2$ and $\epsilon^{(1)}_{4e}=0$ with only considering the following non-zero VEVs,
\begin{align}
 u^d_3  &=  0.00054705, \qquad\quad  u^d_4  =  0.000429559 ,\nonumber\\
 u^d_5  &=  0.01716, \qquad\qquad ~w^d_8  =  1 ,      \nonumber \\
 F^e_1  = F^e_{19} &= 0.595495, \qquad\quad F^e_{17} = 0.00941675 \nonumber\\ 
\Rightarrow ~ Z_{4e} &= m_\tau w^d_8 \left(
\begin{array}{ccc}
 0. & u^d_5 F^e_{17} & u^d_4 F^e_{19} \\
 u^d_5 F^e_{17} & u^d_4 F^e_{19} & u^d_3 F^e_1 \\
 u^d_4 F^e_{19} & u^d_3 F^e_1 & 0. \\
\end{array}
\right), \label{eq:UdWd22}
\end{align}
which yields the following four-point contribution to be added to the 3-point functions $\{Z_{3d},Z_{3e}\}$ \eqref{eq:Leptons3_22} as,
\begin{align}
Z_{4e}&=m_\tau \left(
\begin{array}{ccc}
 0 & 0.156933 & 0.248425 \\
 0.156933 & 0.248425 & 0.316373 \\
 0.248425 & 0.316373 & 0 \\
\end{array}
\right) \nonumber\\
\Rightarrow ~ & Z_{3e}+Z_{4e} = M_e, \qquad Z_{4d} = 0.\label{eq:M4e_22}
\end{align}
\textbf{Conclusion} -- Therefore, we have achieved the precise matching of all fermion masses and mixings from 3-point couplings alone, whereas the 4-point couplings are only needed to account for the leptons' mixings. This constitutes the first precise prediction of Dirac neutrino masses from a consistent string theory setup. The Dirac masses of neutrinos are derived by three-point functions whereas the leptons' mixing need four-point functions which are suppressed by the string-scale $M_{\rm S}$. This is quite satisfactory because the four-point couplings only affect the tiny neutrinos and all other heavier fermions are unaffected by such interactions. An experimental confirmation of the heaviest neutrino-mass at $\sim 50 $~meV will thus validate the model.

The higher-dimensional 4-point operators $\mathcal{W}_4 \supset{1\over {M_{\rm S}}}\Big( Y^{\prime d}_{ijkl} Q_i D^{c}_j H^{\prime d}_k S^L_l + Y^{\prime e}_{ijkl} L_i E^{c}_j H^{\prime d}_k S^L_l \Big) $ needed to explain neutrino-mixings can be related to the dark-dimension scenario \cite{Montero:2022prj} motivated by the emergent strings conjecture \cite{Lee:2019wij}. Dark dimension relates dark matter (5D gravitons), dark energy ($\Lambda$) and axion decay constant ($f_a\lesssim \hat{M}_5$) with the scale of lightest-neutrino ($m_1$). Taking $m_1=6.2$~meV in the relations $\hat{M}_5= m_1^{1/3}M_{\rm pl}^{2/3}$ and $m_{1} = \lambda^{-1} \Lambda^{1/4}$, the species-scale in 5D is set at $\hat{M}_5 =9.74\times 10^8$~GeV resulting in the size and the thickness of the dark-dimension to be 31.8~$\mu$m and $2.0\times 10^{-23}$~cm respectively. No deviations in the gravitational inverse-square law have been detected above $38.6$~$\mu$m at 2$\sigma$ \cite{Lee:2020zjt}, however, it is to be probed in near-future.  

\textbf{Acknowledgments} -- We have benefitted from discussions with Sung-Soo Kim and Miguel Montero. TL is supported in part by the National Key Research and Development Program of China Grant No. 2020YFC2201504, by the Projects No. 11875062, No. 11947302, No. 12047503, and No. 12275333 supported by the National Natural Science Foundation of China, by the Key Research Program of the Chinese Academy of Sciences, Grant No. XDPB15, by the Scientific Instrument Developing Project of the Chinese Academy of Sciences, Grant No. YJKYYQ20190049, and by the International Partnership Program of Chinese Academy of Sciences for Grand Challenges, Grant No. 112311KYSB20210012. AM is supported by the Guangdong Basic and Applied Basic Research Foundation (Grant No. 2021B1515130007), Shenzhen Natural Science Fund (the Stable Support Plan Program 20220810130956001). Z.-W. Wang is supported in part by the hundred talented program at University of Electronic Science and Technology of China and by the National Natural Science Foundation of China (Grant No. 12475105). 

%\bibliographystyle{apsrev4-2}
%\bibliography{References} 

\begin{thebibliography}{47}%
\makeatletter
\providecommand \@ifxundefined [1]{%
 \@ifx{#1\undefined}
}%
\providecommand \@ifnum [1]{%
 \ifnum #1\expandafter \@firstoftwo
 \else \expandafter \@secondoftwo
 \fi
}%
\providecommand \@ifx [1]{%
 \ifx #1\expandafter \@firstoftwo
 \else \expandafter \@secondoftwo
 \fi
}%
\providecommand \natexlab [1]{#1}%
\providecommand \enquote  [1]{``#1''}%
\providecommand \bibnamefont  [1]{#1}%
\providecommand \bibfnamefont [1]{#1}%
\providecommand \citenamefont [1]{#1}%
\providecommand \href@noop [0]{\@secondoftwo}%
\providecommand \href [0]{\begingroup \@sanitize@url \@href}%
\providecommand \@href[1]{\@@startlink{#1}\@@href}%
\providecommand \@@href[1]{\endgroup#1\@@endlink}%
\providecommand \@sanitize@url [0]{\catcode `\\12\catcode `\$12\catcode
  `\&12\catcode `\#12\catcode `\^12\catcode `\_12\catcode `\%12\relax}%
\providecommand \@@startlink[1]{}%
\providecommand \@@endlink[0]{}%
\providecommand \url  [0]{\begingroup\@sanitize@url \@url }%
\providecommand \@url [1]{\endgroup\@href {#1}{\urlprefix }}%
\providecommand \urlprefix  [0]{URL }%
\providecommand \Eprint [0]{\href }%
\providecommand \doibase [0]{https://doi.org/}%
\providecommand \selectlanguage [0]{\@gobble}%
\providecommand \bibinfo  [0]{\@secondoftwo}%
\providecommand \bibfield  [0]{\@secondoftwo}%
\providecommand \translation [1]{[#1]}%
\providecommand \BibitemOpen [0]{}%
\providecommand \bibitemStop [0]{}%
\providecommand \bibitemNoStop [0]{.\EOS\space}%
\providecommand \EOS [0]{\spacefactor3000\relax}%
\providecommand \BibitemShut  [1]{\csname bibitem#1\endcsname}%
\let\auto@bib@innerbib\@empty
%</preamble>
\bibitem [{\citenamefont {Aldazabal}\ \emph {et~al.}(2001)\citenamefont
  {Aldazabal}, \citenamefont {Franco}, \citenamefont {Ibanez}, \citenamefont
  {Rabadan},\ and\ \citenamefont {Uranga}}]{Aldazabal:2000cn}%
  \BibitemOpen
  \bibfield  {author} {\bibinfo {author} {\bibfnamefont {G.}~\bibnamefont
  {Aldazabal}}, \bibinfo {author} {\bibfnamefont {S.}~\bibnamefont {Franco}},
  \bibinfo {author} {\bibfnamefont {L.~E.}\ \bibnamefont {Ibanez}}, \bibinfo
  {author} {\bibfnamefont {R.}~\bibnamefont {Rabadan}},\ and\ \bibinfo {author}
  {\bibfnamefont {A.~M.}\ \bibnamefont {Uranga}},\ }\bibfield  {title}
  {\bibinfo {title} {{Intersecting brane worlds}},\ }\href
  {https://doi.org/10.1088/1126-6708/2001/02/047} {\bibfield  {journal}
  {\bibinfo  {journal} {JHEP}\ }\textbf {\bibinfo {volume} {02}},\ \bibinfo
  {pages} {047}},\ \Eprint {https://arxiv.org/abs/hep-ph/0011132}
  {arXiv:hep-ph/0011132} \BibitemShut {NoStop}%
\bibitem [{\citenamefont {Cremades}\ \emph {et~al.}(2003)\citenamefont
  {Cremades}, \citenamefont {Ibanez},\ and\ \citenamefont
  {Marchesano}}]{Cremades:2003qj}%
  \BibitemOpen
  \bibfield  {author} {\bibinfo {author} {\bibfnamefont {D.}~\bibnamefont
  {Cremades}}, \bibinfo {author} {\bibfnamefont {L.~E.}\ \bibnamefont
  {Ibanez}},\ and\ \bibinfo {author} {\bibfnamefont {F.}~\bibnamefont
  {Marchesano}},\ }\bibfield  {title} {\bibinfo {title} {{Yukawa couplings in
  intersecting D-brane models}},\ }\href
  {https://doi.org/10.1088/1126-6708/2003/07/038} {\bibfield  {journal}
  {\bibinfo  {journal} {JHEP}\ }\textbf {\bibinfo {volume} {07}},\ \bibinfo
  {pages} {038}},\ \Eprint {https://arxiv.org/abs/hep-th/0302105}
  {arXiv:hep-th/0302105} \BibitemShut {NoStop}%
\bibitem [{\citenamefont {Witten}(1998)}]{Witten:1998cd}%
  \BibitemOpen
  \bibfield  {author} {\bibinfo {author} {\bibfnamefont {E.}~\bibnamefont
  {Witten}},\ }\bibfield  {title} {\bibinfo {title} {{D-branes and K-theory}},\
  }\href {https://doi.org/10.1088/1126-6708/1998/12/019} {\bibfield  {journal}
  {\bibinfo  {journal} {JHEP}\ }\textbf {\bibinfo {volume} {12}},\ \bibinfo
  {pages} {019}},\ \Eprint {https://arxiv.org/abs/hep-th/9810188}
  {arXiv:hep-th/9810188} \BibitemShut {NoStop}%
\bibitem [{\citenamefont {Uranga}(2001)}]{Uranga:2000xp}%
  \BibitemOpen
  \bibfield  {author} {\bibinfo {author} {\bibfnamefont {A.~M.}\ \bibnamefont
  {Uranga}},\ }\bibfield  {title} {\bibinfo {title} {{D-brane probes, RR
  tadpole cancellation and K-theory charge}},\ }\href
  {https://doi.org/10.1016/S0550-3213(00)00787-2} {\bibfield  {journal}
  {\bibinfo  {journal} {Nucl. Phys. B}\ }\textbf {\bibinfo {volume} {598}},\
  \bibinfo {pages} {225} (\bibinfo {year} {2001})},\ \Eprint
  {https://arxiv.org/abs/hep-th/0011048} {arXiv:hep-th/0011048} \BibitemShut
  {NoStop}%
\bibitem [{\citenamefont {Cvetic}\ \emph {et~al.}(2004)\citenamefont {Cvetic},
  \citenamefont {Li},\ and\ \citenamefont {Liu}}]{Cvetic:2004ui}%
  \BibitemOpen
  \bibfield  {author} {\bibinfo {author} {\bibfnamefont {M.}~\bibnamefont
  {Cvetic}}, \bibinfo {author} {\bibfnamefont {T.}~\bibnamefont {Li}},\ and\
  \bibinfo {author} {\bibfnamefont {T.}~\bibnamefont {Liu}},\ }\bibfield
  {title} {\bibinfo {title} {{Supersymmetric Pati-Salam models from
  intersecting D6-branes: A Road to the standard model}},\ }\href
  {https://doi.org/10.1016/j.nuclphysb.2004.07.036} {\bibfield  {journal}
  {\bibinfo  {journal} {Nucl. Phys. B}\ }\textbf {\bibinfo {volume} {698}},\
  \bibinfo {pages} {163} (\bibinfo {year} {2004})},\ \Eprint
  {https://arxiv.org/abs/hep-th/0403061} {arXiv:hep-th/0403061} \BibitemShut
  {NoStop}%
\bibitem [{\citenamefont {Blumenhagen}\ \emph
  {et~al.}(2007{\natexlab{a}})\citenamefont {Blumenhagen}, \citenamefont
  {Kors}, \citenamefont {Lust},\ and\ \citenamefont
  {Stieberger}}]{Blumenhagen:2006ci}%
  \BibitemOpen
  \bibfield  {author} {\bibinfo {author} {\bibfnamefont {R.}~\bibnamefont
  {Blumenhagen}}, \bibinfo {author} {\bibfnamefont {B.}~\bibnamefont {Kors}},
  \bibinfo {author} {\bibfnamefont {D.}~\bibnamefont {Lust}},\ and\ \bibinfo
  {author} {\bibfnamefont {S.}~\bibnamefont {Stieberger}},\ }\bibfield  {title}
  {\bibinfo {title} {{Four-dimensional String Compactifications with D-Branes,
  Orientifolds and Fluxes}},\ }\href
  {https://doi.org/10.1016/j.physrep.2007.04.003} {\bibfield  {journal}
  {\bibinfo  {journal} {Phys. Rept.}\ }\textbf {\bibinfo {volume} {445}},\
  \bibinfo {pages} {1} (\bibinfo {year} {2007}{\natexlab{a}})},\ \Eprint
  {https://arxiv.org/abs/hep-th/0610327} {arXiv:hep-th/0610327} \BibitemShut
  {NoStop}%
\bibitem [{\citenamefont {Blumenhagen}\ \emph {et~al.}(2005)\citenamefont
  {Blumenhagen}, \citenamefont {Cvetic}, \citenamefont {Langacker},\ and\
  \citenamefont {Shiu}}]{Blumenhagen:2005mu}%
  \BibitemOpen
  \bibfield  {author} {\bibinfo {author} {\bibfnamefont {R.}~\bibnamefont
  {Blumenhagen}}, \bibinfo {author} {\bibfnamefont {M.}~\bibnamefont {Cvetic}},
  \bibinfo {author} {\bibfnamefont {P.}~\bibnamefont {Langacker}},\ and\
  \bibinfo {author} {\bibfnamefont {G.}~\bibnamefont {Shiu}},\ }\bibfield
  {title} {\bibinfo {title} {{Toward realistic intersecting D-brane models}},\
  }\href {https://doi.org/10.1146/annurev.nucl.55.090704.151541} {\bibfield
  {journal} {\bibinfo  {journal} {Ann. Rev. Nucl. Part. Sci.}\ }\textbf
  {\bibinfo {volume} {55}},\ \bibinfo {pages} {71} (\bibinfo {year} {2005})},\
  \Eprint {https://arxiv.org/abs/hep-th/0502005} {arXiv:hep-th/0502005}
  \BibitemShut {NoStop}%
\bibitem [{\citenamefont {Li}\ \emph {et~al.}(2021{\natexlab{a}})\citenamefont
  {Li}, \citenamefont {Mansha},\ and\ \citenamefont {Sun}}]{Li:2019nvi}%
  \BibitemOpen
  \bibfield  {author} {\bibinfo {author} {\bibfnamefont {T.}~\bibnamefont
  {Li}}, \bibinfo {author} {\bibfnamefont {A.}~\bibnamefont {Mansha}},\ and\
  \bibinfo {author} {\bibfnamefont {R.}~\bibnamefont {Sun}},\ }\bibfield
  {title} {\bibinfo {title} {{Revisiting the supersymmetric
  Pati\textendash{}Salam models from intersecting D6-branes}},\ }\href
  {https://doi.org/10.1140/epjc/s10052-021-08839-w} {\bibfield  {journal}
  {\bibinfo  {journal} {Eur. Phys. J. C}\ }\textbf {\bibinfo {volume} {81}},\
  \bibinfo {pages} {82} (\bibinfo {year} {2021}{\natexlab{a}})},\ \Eprint
  {https://arxiv.org/abs/1910.04530} {arXiv:1910.04530 [hep-th]} \BibitemShut
  {NoStop}%
\bibitem [{\citenamefont {Li}\ \emph {et~al.}(2021{\natexlab{b}})\citenamefont
  {Li}, \citenamefont {Mansha}, \citenamefont {Sun}, \citenamefont {Wu},\ and\
  \citenamefont {He}}]{Li:2021pxo}%
  \BibitemOpen
  \bibfield  {author} {\bibinfo {author} {\bibfnamefont {T.}~\bibnamefont
  {Li}}, \bibinfo {author} {\bibfnamefont {A.}~\bibnamefont {Mansha}}, \bibinfo
  {author} {\bibfnamefont {R.}~\bibnamefont {Sun}}, \bibinfo {author}
  {\bibfnamefont {L.}~\bibnamefont {Wu}},\ and\ \bibinfo {author}
  {\bibfnamefont {W.}~\bibnamefont {He}},\ }\bibfield  {title} {\bibinfo
  {title} {{N=1 supersymmetric $SU(12)_C \times SU (2)_L \times SU(2)_R$
  models,~$SU(4)_C \times SU(6)_L \times SU(2)_R$~models, and $SU(4)_C \times
  SU(2)_L \times SU(6)_R$~ models from intersecting D6-branes}},\ }\href
  {https://doi.org/10.1103/PhysRevD.104.046018} {\bibfield  {journal} {\bibinfo
   {journal} {Phys. Rev. D}\ }\textbf {\bibinfo {volume} {104}},\ \bibinfo
  {pages} {046018} (\bibinfo {year} {2021}{\natexlab{b}})}\BibitemShut
  {NoStop}%
\bibitem [{\citenamefont {Mansha}\ \emph
  {et~al.}(2024{\natexlab{a}})\citenamefont {Mansha}, \citenamefont {Li},
  \citenamefont {Sabir},\ and\ \citenamefont {Wu}}]{Mansha:2022pnd}%
  \BibitemOpen
  \bibfield  {author} {\bibinfo {author} {\bibfnamefont {A.}~\bibnamefont
  {Mansha}}, \bibinfo {author} {\bibfnamefont {T.}~\bibnamefont {Li}}, \bibinfo
  {author} {\bibfnamefont {M.}~\bibnamefont {Sabir}},\ and\ \bibinfo {author}
  {\bibfnamefont {L.}~\bibnamefont {Wu}},\ }\bibfield  {title} {\bibinfo
  {title} {{Three-family supersymmetric Pati\textendash{}Salam models with
  symplectic groups from intersecting D6-branes}},\ }\href
  {https://doi.org/10.1140/epjc/s10052-024-12411-7} {\bibfield  {journal}
  {\bibinfo  {journal} {Eur. Phys. J. C}\ }\textbf {\bibinfo {volume} {84}},\
  \bibinfo {pages} {151} (\bibinfo {year} {2024}{\natexlab{a}})},\ \Eprint
  {https://arxiv.org/abs/2212.09644} {arXiv:2212.09644 [hep-th]} \BibitemShut
  {NoStop}%
\bibitem [{\citenamefont {Sabir}\ \emph {et~al.}(2022)\citenamefont {Sabir},
  \citenamefont {Li}, \citenamefont {Mansha},\ and\ \citenamefont
  {Wang}}]{Sabir:2022hko}%
  \BibitemOpen
  \bibfield  {author} {\bibinfo {author} {\bibfnamefont {M.}~\bibnamefont
  {Sabir}}, \bibinfo {author} {\bibfnamefont {T.}~\bibnamefont {Li}}, \bibinfo
  {author} {\bibfnamefont {A.}~\bibnamefont {Mansha}},\ and\ \bibinfo {author}
  {\bibfnamefont {X.-C.}\ \bibnamefont {Wang}},\ }\bibfield  {title} {\bibinfo
  {title} {{The supersymmetry breaking soft terms, and fermion masses and
  mixings in the supersymmetric Pati-Salam model from intersecting
  D6-branes}},\ }\href {https://doi.org/10.1007/JHEP04(2022)089} {\bibfield
  {journal} {\bibinfo  {journal} {JHEP}\ }\textbf {\bibinfo {volume} {04}},\
  \bibinfo {pages} {089}},\ \Eprint {https://arxiv.org/abs/2202.07048}
  {arXiv:2202.07048 [hep-th]} \BibitemShut {NoStop}%
\bibitem [{\citenamefont {Mansha}\ \emph {et~al.}(2023)\citenamefont {Mansha},
  \citenamefont {Li},\ and\ \citenamefont {Wu}}]{Mansha:2023kwq}%
  \BibitemOpen
  \bibfield  {author} {\bibinfo {author} {\bibfnamefont {A.}~\bibnamefont
  {Mansha}}, \bibinfo {author} {\bibfnamefont {T.}~\bibnamefont {Li}},\ and\
  \bibinfo {author} {\bibfnamefont {L.}~\bibnamefont {Wu}},\ }\bibfield
  {title} {\bibinfo {title} {{The hidden sector variations in the
  $\mathcal{N}=1$ supersymmetric three-family Pati\textendash{}Salam models
  from intersecting D6-branes}},\ }\href
  {https://doi.org/10.1140/epjc/s10052-023-12167-6} {\bibfield  {journal}
  {\bibinfo  {journal} {Eur. Phys. J. C}\ }\textbf {\bibinfo {volume} {83}},\
  \bibinfo {pages} {1067} (\bibinfo {year} {2023})},\ \Eprint
  {https://arxiv.org/abs/2303.02864} {arXiv:2303.02864 [hep-th]} \BibitemShut
  {NoStop}%
\bibitem [{\citenamefont {Mansha}\ \emph
  {et~al.}(2024{\natexlab{b}})\citenamefont {Mansha}, \citenamefont {Li},\ and\
  \citenamefont {Sabir}}]{Mansha:2024yqz}%
  \BibitemOpen
  \bibfield  {author} {\bibinfo {author} {\bibfnamefont {A.}~\bibnamefont
  {Mansha}}, \bibinfo {author} {\bibfnamefont {T.}~\bibnamefont {Li}},\ and\
  \bibinfo {author} {\bibfnamefont {M.}~\bibnamefont {Sabir}},\ }\bibfield
  {title} {\bibinfo {title} {{Revisiting the supersymmetric trinification
  models from intersecting D6-branes}},\ }\href
  {https://doi.org/10.1088/1572-9494/ad565f} {\bibfield  {journal} {\bibinfo
  {journal} {Commun. Theor. Phys.}\ }\textbf {\bibinfo {volume} {76}},\
  \bibinfo {pages} {095201} (\bibinfo {year} {2024}{\natexlab{b}})},\ \Eprint
  {https://arxiv.org/abs/2406.07586} {arXiv:2406.07586 [hep-th]} \BibitemShut
  {NoStop}%
\bibitem [{\citenamefont {He}\ \emph {et~al.}(2022)\citenamefont {He},
  \citenamefont {Li},\ and\ \citenamefont {Sun}}]{He:2021gug}%
  \BibitemOpen
  \bibfield  {author} {\bibinfo {author} {\bibfnamefont {W.}~\bibnamefont
  {He}}, \bibinfo {author} {\bibfnamefont {T.}~\bibnamefont {Li}},\ and\
  \bibinfo {author} {\bibfnamefont {R.}~\bibnamefont {Sun}},\ }\bibfield
  {title} {\bibinfo {title} {{The complete search for the supersymmetric
  Pati-Salam models from intersecting D6-branes}},\ }\href
  {https://doi.org/10.1007/JHEP08(2022)044} {\bibfield  {journal} {\bibinfo
  {journal} {JHEP}\ }\textbf {\bibinfo {volume} {08}},\ \bibinfo {pages}
  {044}},\ \Eprint {https://arxiv.org/abs/2112.09632} {arXiv:2112.09632
  [hep-th]} \BibitemShut {NoStop}%
\bibitem [{\citenamefont {Lee}\ \emph {et~al.}(2022)\citenamefont {Lee},
  \citenamefont {Lerche},\ and\ \citenamefont {Weigand}}]{Lee:2019wij}%
  \BibitemOpen
  \bibfield  {author} {\bibinfo {author} {\bibfnamefont {S.-J.}\ \bibnamefont
  {Lee}}, \bibinfo {author} {\bibfnamefont {W.}~\bibnamefont {Lerche}},\ and\
  \bibinfo {author} {\bibfnamefont {T.}~\bibnamefont {Weigand}},\ }\bibfield
  {title} {\bibinfo {title} {{Emergent strings from infinite distance
  limits}},\ }\href {https://doi.org/10.1007/JHEP02(2022)190} {\bibfield
  {journal} {\bibinfo  {journal} {JHEP}\ }\textbf {\bibinfo {volume} {02}},\
  \bibinfo {pages} {190}},\ \Eprint {https://arxiv.org/abs/1910.01135}
  {arXiv:1910.01135 [hep-th]} \BibitemShut {NoStop}%
\bibitem [{\citenamefont {Casas}\ \emph {et~al.}(2024)\citenamefont {Casas},
  \citenamefont {Ib\'a\~nez},\ and\ \citenamefont
  {Marchesano}}]{Casas:2024ttx}%
  \BibitemOpen
  \bibfield  {author} {\bibinfo {author} {\bibfnamefont {G.~F.}\ \bibnamefont
  {Casas}}, \bibinfo {author} {\bibfnamefont {L.~E.}\ \bibnamefont
  {Ib\'a\~nez}},\ and\ \bibinfo {author} {\bibfnamefont {F.}~\bibnamefont
  {Marchesano}},\ }\bibfield  {title} {\bibinfo {title} {{Yukawa couplings at
  infinite distance and swampland towers in chiral theories}},\ }\href
  {https://doi.org/10.1007/JHEP09(2024)170} {\bibfield  {journal} {\bibinfo
  {journal} {JHEP}\ }\textbf {\bibinfo {volume} {09}},\ \bibinfo {pages}
  {170}},\ \Eprint {https://arxiv.org/abs/2403.09775} {arXiv:2403.09775
  [hep-th]} \BibitemShut {NoStop}%
\bibitem [{\citenamefont {Casas}\ \emph {et~al.}(2025)\citenamefont {Casas},
  \citenamefont {Ib\'a\~nez},\ and\ \citenamefont
  {Marchesano}}]{Casas:2024clw}%
  \BibitemOpen
  \bibfield  {author} {\bibinfo {author} {\bibfnamefont {G.~F.}\ \bibnamefont
  {Casas}}, \bibinfo {author} {\bibfnamefont {L.~E.}\ \bibnamefont
  {Ib\'a\~nez}},\ and\ \bibinfo {author} {\bibfnamefont {F.}~\bibnamefont
  {Marchesano}},\ }\bibfield  {title} {\bibinfo {title} {{On small Dirac
  neutrino masses in string theory}},\ }\href
  {https://doi.org/10.1007/JHEP01(2025)083} {\bibfield  {journal} {\bibinfo
  {journal} {JHEP}\ }\textbf {\bibinfo {volume} {01}},\ \bibinfo {pages}
  {083}},\ \Eprint {https://arxiv.org/abs/2406.14609} {arXiv:2406.14609
  [hep-th]} \BibitemShut {NoStop}%
\bibitem [{\citenamefont {Sabir}\ \emph {et~al.}(2024)\citenamefont {Sabir},
  \citenamefont {Mansha}, \citenamefont {Li},\ and\ \citenamefont
  {Wang}}]{Sabir:2024cgt}%
  \BibitemOpen
  \bibfield  {author} {\bibinfo {author} {\bibfnamefont {M.}~\bibnamefont
  {Sabir}}, \bibinfo {author} {\bibfnamefont {A.}~\bibnamefont {Mansha}},
  \bibinfo {author} {\bibfnamefont {T.}~\bibnamefont {Li}},\ and\ \bibinfo
  {author} {\bibfnamefont {Z.-W.}\ \bibnamefont {Wang}},\ }\bibfield  {title}
  {\bibinfo {title} {{Fermion masses and mixings in the supersymmetric
  Pati-Salam landscape from Intersecting D6-Branes}},\ }\href
  {https://doi.org/10.1007/JHEP10(2024)252} {\bibfield  {journal} {\bibinfo
  {journal} {JHEP}\ }\textbf {\bibinfo {volume} {10}},\ \bibinfo {pages}
  {252}},\ \Eprint {https://arxiv.org/abs/2409.09110} {arXiv:2409.09110
  [hep-ph]} \BibitemShut {NoStop}%
\bibitem [{\citenamefont {Sabir}\ \emph {et~al.}(2025)\citenamefont {Sabir},
  \citenamefont {Mansha}, \citenamefont {Li},\ and\ \citenamefont
  {Wang}}]{Sabir:2024jsx}%
  \BibitemOpen
  \bibfield  {author} {\bibinfo {author} {\bibfnamefont {M.}~\bibnamefont
  {Sabir}}, \bibinfo {author} {\bibfnamefont {A.}~\bibnamefont {Mansha}},
  \bibinfo {author} {\bibfnamefont {T.}~\bibnamefont {Li}},\ and\ \bibinfo
  {author} {\bibfnamefont {Z.-W.}\ \bibnamefont {Wang}},\ }\bibfield  {title}
  {\bibinfo {title} {{Susy breaking soft terms in the supersymmetric Pati-Salam
  landscape from intersecting D6-branes}},\ }\href
  {https://doi.org/10.1007/JHEP01(2025)201} {\bibfield  {journal} {\bibinfo
  {journal} {JHEP}\ }\textbf {\bibinfo {volume} {01}},\ \bibinfo {pages}
  {201}},\ \Eprint {https://arxiv.org/abs/2410.09093} {arXiv:2410.09093
  [hep-ph]} \BibitemShut {NoStop}%
\bibitem [{\citenamefont {Esteban}\ \emph {et~al.}(2024)\citenamefont
  {Esteban}, \citenamefont {Gonzalez-Garcia}, \citenamefont {Maltoni},
  \citenamefont {Martinez-Soler}, \citenamefont {Pinheiro},\ and\ \citenamefont
  {Schwetz}}]{Esteban:2024eli}%
  \BibitemOpen
  \bibfield  {author} {\bibinfo {author} {\bibfnamefont {I.}~\bibnamefont
  {Esteban}}, \bibinfo {author} {\bibfnamefont {M.~C.}\ \bibnamefont
  {Gonzalez-Garcia}}, \bibinfo {author} {\bibfnamefont {M.}~\bibnamefont
  {Maltoni}}, \bibinfo {author} {\bibfnamefont {I.}~\bibnamefont
  {Martinez-Soler}}, \bibinfo {author} {\bibfnamefont {J.~a.~P.}\ \bibnamefont
  {Pinheiro}},\ and\ \bibinfo {author} {\bibfnamefont {T.}~\bibnamefont
  {Schwetz}},\ }\bibfield  {title} {\bibinfo {title} {{NuFit-6.0: updated
  global analysis of three-flavor neutrino oscillations}},\ }\href
  {https://doi.org/10.1007/JHEP12(2024)216} {\bibfield  {journal} {\bibinfo
  {journal} {JHEP}\ }\textbf {\bibinfo {volume} {12}},\ \bibinfo {pages}
  {216}},\ \Eprint {https://arxiv.org/abs/2410.05380} {arXiv:2410.05380
  [hep-ph]} \BibitemShut {NoStop}%
\bibitem [{\citenamefont {Adame}\ \emph {et~al.}(2025)\citenamefont {Adame}
  \emph {et~al.}}]{DESI:2024mwx}%
  \BibitemOpen
  \bibfield  {author} {\bibinfo {author} {\bibfnamefont {A.~G.}\ \bibnamefont
  {Adame}} \emph {et~al.} (\bibinfo {collaboration} {DESI}),\ }\bibfield
  {title} {\bibinfo {title} {{DESI 2024 VI: cosmological constraints from the
  measurements of baryon acoustic oscillations}},\ }\href
  {https://doi.org/10.1088/1475-7516/2025/02/021} {\bibfield  {journal}
  {\bibinfo  {journal} {JCAP}\ }\textbf {\bibinfo {volume} {02}},\ \bibinfo
  {pages} {021}},\ \Eprint {https://arxiv.org/abs/2404.03002} {arXiv:2404.03002
  [astro-ph.CO]} \BibitemShut {NoStop}%
\bibitem [{\citenamefont {Ooguri}\ and\ \citenamefont
  {Vafa}(2017)}]{Ooguri:2016pdq}%
  \BibitemOpen
  \bibfield  {author} {\bibinfo {author} {\bibfnamefont {H.}~\bibnamefont
  {Ooguri}}\ and\ \bibinfo {author} {\bibfnamefont {C.}~\bibnamefont {Vafa}},\
  }\bibfield  {title} {\bibinfo {title} {{Non-supersymmetric AdS and the
  Swampland}},\ }\href {https://doi.org/10.4310/ATMP.2017.v21.n7.a8} {\bibfield
   {journal} {\bibinfo  {journal} {Adv. Theor. Math. Phys.}\ }\textbf {\bibinfo
  {volume} {21}},\ \bibinfo {pages} {1787} (\bibinfo {year} {2017})},\ \Eprint
  {https://arxiv.org/abs/1610.01533} {arXiv:1610.01533 [hep-th]} \BibitemShut
  {NoStop}%
\bibitem [{\citenamefont {Gonzalo}\ \emph {et~al.}(2021)\citenamefont
  {Gonzalo}, \citenamefont {Ib\'a\~nez},\ and\ \citenamefont
  {Valenzuela}}]{Gonzalo:2021fma}%
  \BibitemOpen
  \bibfield  {author} {\bibinfo {author} {\bibfnamefont {E.}~\bibnamefont
  {Gonzalo}}, \bibinfo {author} {\bibfnamefont {L.~E.}\ \bibnamefont
  {Ib\'a\~nez}},\ and\ \bibinfo {author} {\bibfnamefont {I.}~\bibnamefont
  {Valenzuela}},\ }\bibfield  {title} {\bibinfo {title} {{AdS swampland
  conjectures and light fermions}},\ }\href
  {https://doi.org/10.1016/j.physletb.2021.136691} {\bibfield  {journal}
  {\bibinfo  {journal} {Phys. Lett. B}\ }\textbf {\bibinfo {volume} {822}},\
  \bibinfo {pages} {136691} (\bibinfo {year} {2021})},\ \Eprint
  {https://arxiv.org/abs/2104.06415} {arXiv:2104.06415 [hep-th]} \BibitemShut
  {NoStop}%
\bibitem [{\citenamefont {Arkani-Hamed}\ \emph {et~al.}(2007)\citenamefont
  {Arkani-Hamed}, \citenamefont {Dubovsky}, \citenamefont {Nicolis},\ and\
  \citenamefont {Villadoro}}]{Arkani-Hamed:2007ryu}%
  \BibitemOpen
  \bibfield  {author} {\bibinfo {author} {\bibfnamefont {N.}~\bibnamefont
  {Arkani-Hamed}}, \bibinfo {author} {\bibfnamefont {S.}~\bibnamefont
  {Dubovsky}}, \bibinfo {author} {\bibfnamefont {A.}~\bibnamefont {Nicolis}},\
  and\ \bibinfo {author} {\bibfnamefont {G.}~\bibnamefont {Villadoro}},\
  }\bibfield  {title} {\bibinfo {title} {{Quantum Horizons of the Standard
  Model Landscape}},\ }\href {https://doi.org/10.1088/1126-6708/2007/06/078}
  {\bibfield  {journal} {\bibinfo  {journal} {JHEP}\ }\textbf {\bibinfo
  {volume} {06}},\ \bibinfo {pages} {078}},\ \Eprint
  {https://arxiv.org/abs/hep-th/0703067} {arXiv:hep-th/0703067} \BibitemShut
  {NoStop}%
\bibitem [{\citenamefont {Arnold}\ \emph {et~al.}(2010)\citenamefont {Arnold},
  \citenamefont {Fornal},\ and\ \citenamefont {Wise}}]{Arnold:2010qz}%
  \BibitemOpen
  \bibfield  {author} {\bibinfo {author} {\bibfnamefont {J.~M.}\ \bibnamefont
  {Arnold}}, \bibinfo {author} {\bibfnamefont {B.}~\bibnamefont {Fornal}},\
  and\ \bibinfo {author} {\bibfnamefont {M.~B.}\ \bibnamefont {Wise}},\
  }\bibfield  {title} {\bibinfo {title} {{Standard Model Vacua for
  Two-dimensional Compactifications}},\ }\href
  {https://doi.org/10.1007/JHEP12(2010)083} {\bibfield  {journal} {\bibinfo
  {journal} {JHEP}\ }\textbf {\bibinfo {volume} {12}},\ \bibinfo {pages}
  {083}},\ \Eprint {https://arxiv.org/abs/1010.4302} {arXiv:1010.4302 [hep-th]}
  \BibitemShut {NoStop}%
\bibitem [{\citenamefont {Ibanez}\ \emph {et~al.}(2017)\citenamefont {Ibanez},
  \citenamefont {Martin-Lozano},\ and\ \citenamefont
  {Valenzuela}}]{Ibanez:2017kvh}%
  \BibitemOpen
  \bibfield  {author} {\bibinfo {author} {\bibfnamefont {L.~E.}\ \bibnamefont
  {Ibanez}}, \bibinfo {author} {\bibfnamefont {V.}~\bibnamefont
  {Martin-Lozano}},\ and\ \bibinfo {author} {\bibfnamefont {I.}~\bibnamefont
  {Valenzuela}},\ }\bibfield  {title} {\bibinfo {title} {{Constraining Neutrino
  Masses, the Cosmological Constant and BSM Physics from the Weak Gravity
  Conjecture}},\ }\href {https://doi.org/10.1007/JHEP11(2017)066} {\bibfield
  {journal} {\bibinfo  {journal} {JHEP}\ }\textbf {\bibinfo {volume} {11}},\
  \bibinfo {pages} {066}},\ \Eprint {https://arxiv.org/abs/1706.05392}
  {arXiv:1706.05392 [hep-th]} \BibitemShut {NoStop}%
\bibitem [{\citenamefont {Hamada}\ and\ \citenamefont
  {Shiu}(2017)}]{Hamada:2017yji}%
  \BibitemOpen
  \bibfield  {author} {\bibinfo {author} {\bibfnamefont {Y.}~\bibnamefont
  {Hamada}}\ and\ \bibinfo {author} {\bibfnamefont {G.}~\bibnamefont {Shiu}},\
  }\bibfield  {title} {\bibinfo {title} {{Weak Gravity Conjecture, Multiple
  Point Principle and the Standard Model Landscape}},\ }\href
  {https://doi.org/10.1007/JHEP11(2017)043} {\bibfield  {journal} {\bibinfo
  {journal} {JHEP}\ }\textbf {\bibinfo {volume} {11}},\ \bibinfo {pages}
  {043}},\ \Eprint {https://arxiv.org/abs/1707.06326} {arXiv:1707.06326
  [hep-th]} \BibitemShut {NoStop}%
\bibitem [{\citenamefont {Gonzalo}\ \emph {et~al.}(2018)\citenamefont
  {Gonzalo}, \citenamefont {Herr\'aez},\ and\ \citenamefont
  {Ib\'a\~nez}}]{Gonzalo:2018tpb}%
  \BibitemOpen
  \bibfield  {author} {\bibinfo {author} {\bibfnamefont {E.}~\bibnamefont
  {Gonzalo}}, \bibinfo {author} {\bibfnamefont {A.}~\bibnamefont {Herr\'aez}},\
  and\ \bibinfo {author} {\bibfnamefont {L.~E.}\ \bibnamefont {Ib\'a\~nez}},\
  }\bibfield  {title} {\bibinfo {title} {{AdS-phobia, the WGC, the Standard
  Model and Supersymmetry}},\ }\href {https://doi.org/10.1007/JHEP06(2018)051}
  {\bibfield  {journal} {\bibinfo  {journal} {JHEP}\ }\textbf {\bibinfo
  {volume} {06}},\ \bibinfo {pages} {051}},\ \Eprint
  {https://arxiv.org/abs/1803.08455} {arXiv:1803.08455 [hep-th]} \BibitemShut
  {NoStop}%
\bibitem [{\citenamefont {Gonzalo}\ \emph {et~al.}(2022)\citenamefont
  {Gonzalo}, \citenamefont {Ib\'a\~nez},\ and\ \citenamefont
  {Valenzuela}}]{Gonzalo:2021zsp}%
  \BibitemOpen
  \bibfield  {author} {\bibinfo {author} {\bibfnamefont {E.}~\bibnamefont
  {Gonzalo}}, \bibinfo {author} {\bibfnamefont {L.~E.}\ \bibnamefont
  {Ib\'a\~nez}},\ and\ \bibinfo {author} {\bibfnamefont {I.}~\bibnamefont
  {Valenzuela}},\ }\bibfield  {title} {\bibinfo {title} {{Swampland constraints
  on neutrino masses}},\ }\href {https://doi.org/10.1007/JHEP02(2022)088}
  {\bibfield  {journal} {\bibinfo  {journal} {JHEP}\ }\textbf {\bibinfo
  {volume} {02}},\ \bibinfo {pages} {088}},\ \Eprint
  {https://arxiv.org/abs/2109.10961} {arXiv:2109.10961 [hep-th]} \BibitemShut
  {NoStop}%
\bibitem [{\citenamefont {Castellano}\ \emph {et~al.}(2023)\citenamefont
  {Castellano}, \citenamefont {Herr\'aez},\ and\ \citenamefont
  {Ib\'a\~nez}}]{Castellano:2023qhp}%
  \BibitemOpen
  \bibfield  {author} {\bibinfo {author} {\bibfnamefont {A.}~\bibnamefont
  {Castellano}}, \bibinfo {author} {\bibfnamefont {A.}~\bibnamefont
  {Herr\'aez}},\ and\ \bibinfo {author} {\bibfnamefont {L.~E.}\ \bibnamefont
  {Ib\'a\~nez}},\ }\bibfield  {title} {\bibinfo {title} {{Towers and
  hierarchies in the Standard Model from Emergence in Quantum Gravity}},\
  }\href {https://doi.org/10.1007/JHEP10(2023)172} {\bibfield  {journal}
  {\bibinfo  {journal} {JHEP}\ }\textbf {\bibinfo {volume} {10}},\ \bibinfo
  {pages} {172}},\ \Eprint {https://arxiv.org/abs/2302.00017} {arXiv:2302.00017
  [hep-th]} \BibitemShut {NoStop}%
\bibitem [{\citenamefont {Anchordoqui}\ \emph {et~al.}(2024)\citenamefont
  {Anchordoqui}, \citenamefont {Antoniadis},\ and\ \citenamefont
  {Cunat}}]{Anchordoqui:2023wkm}%
  \BibitemOpen
  \bibfield  {author} {\bibinfo {author} {\bibfnamefont {L.~A.}\ \bibnamefont
  {Anchordoqui}}, \bibinfo {author} {\bibfnamefont {I.}~\bibnamefont
  {Antoniadis}},\ and\ \bibinfo {author} {\bibfnamefont {J.}~\bibnamefont
  {Cunat}},\ }\bibfield  {title} {\bibinfo {title} {{Dark dimension and the
  standard model landscape}},\ }\href
  {https://doi.org/10.1103/PhysRevD.109.016028} {\bibfield  {journal} {\bibinfo
   {journal} {Phys. Rev. D}\ }\textbf {\bibinfo {volume} {109}},\ \bibinfo
  {pages} {016028} (\bibinfo {year} {2024})},\ \Eprint
  {https://arxiv.org/abs/2306.16491} {arXiv:2306.16491 [hep-ph]} \BibitemShut
  {NoStop}%
\bibitem [{\citenamefont {Froggatt}\ and\ \citenamefont
  {Nielsen}(1996)}]{Froggatt:1995rt}%
  \BibitemOpen
  \bibfield  {author} {\bibinfo {author} {\bibfnamefont {C.~D.}\ \bibnamefont
  {Froggatt}}\ and\ \bibinfo {author} {\bibfnamefont {H.~B.}\ \bibnamefont
  {Nielsen}},\ }\bibfield  {title} {\bibinfo {title} {{Standard model
  criticality prediction: Top mass 173 +- 5-GeV and Higgs mass 135 +- 9-GeV}},\
  }\href {https://doi.org/10.1016/0370-2693(95)01480-2} {\bibfield  {journal}
  {\bibinfo  {journal} {Phys. Lett. B}\ }\textbf {\bibinfo {volume} {368}},\
  \bibinfo {pages} {96} (\bibinfo {year} {1996})},\ \Eprint
  {https://arxiv.org/abs/hep-ph/9511371} {arXiv:hep-ph/9511371} \BibitemShut
  {NoStop}%
\bibitem [{\citenamefont {L\"ust}\ \emph {et~al.}(2019)\citenamefont {L\"ust},
  \citenamefont {Palti},\ and\ \citenamefont {Vafa}}]{Lust:2019zwm}%
  \BibitemOpen
  \bibfield  {author} {\bibinfo {author} {\bibfnamefont {D.}~\bibnamefont
  {L\"ust}}, \bibinfo {author} {\bibfnamefont {E.}~\bibnamefont {Palti}},\ and\
  \bibinfo {author} {\bibfnamefont {C.}~\bibnamefont {Vafa}},\ }\bibfield
  {title} {\bibinfo {title} {{AdS and the Swampland}},\ }\href
  {https://doi.org/10.1016/j.physletb.2019.134867} {\bibfield  {journal}
  {\bibinfo  {journal} {Phys. Lett. B}\ }\textbf {\bibinfo {volume} {797}},\
  \bibinfo {pages} {134867} (\bibinfo {year} {2019})},\ \Eprint
  {https://arxiv.org/abs/1906.05225} {arXiv:1906.05225 [hep-th]} \BibitemShut
  {NoStop}%
\bibitem [{\citenamefont {Blumenhagen}\ \emph {et~al.}(2009)\citenamefont
  {Blumenhagen}, \citenamefont {Cvetic}, \citenamefont {Kachru},\ and\
  \citenamefont {Weigand}}]{Blumenhagen:2009qh}%
  \BibitemOpen
  \bibfield  {author} {\bibinfo {author} {\bibfnamefont {R.}~\bibnamefont
  {Blumenhagen}}, \bibinfo {author} {\bibfnamefont {M.}~\bibnamefont {Cvetic}},
  \bibinfo {author} {\bibfnamefont {S.}~\bibnamefont {Kachru}},\ and\ \bibinfo
  {author} {\bibfnamefont {T.}~\bibnamefont {Weigand}},\ }\bibfield  {title}
  {\bibinfo {title} {{D-Brane Instantons in Type II Orientifolds}},\ }\href
  {https://doi.org/10.1146/annurev.nucl.010909.083113} {\bibfield  {journal}
  {\bibinfo  {journal} {Ann. Rev. Nucl. Part. Sci.}\ }\textbf {\bibinfo
  {volume} {59}},\ \bibinfo {pages} {269} (\bibinfo {year} {2009})},\ \Eprint
  {https://arxiv.org/abs/0902.3251} {arXiv:0902.3251 [hep-th]} \BibitemShut
  {NoStop}%
\bibitem [{\citenamefont {Cvetic}\ and\ \citenamefont
  {Langacker}(2008)}]{Cvetic:2008hi}%
  \BibitemOpen
  \bibfield  {author} {\bibinfo {author} {\bibfnamefont {M.}~\bibnamefont
  {Cvetic}}\ and\ \bibinfo {author} {\bibfnamefont {P.}~\bibnamefont
  {Langacker}},\ }\bibfield  {title} {\bibinfo {title} {{D-Instanton Generated
  Dirac Neutrino Masses}},\ }\href {https://doi.org/10.1103/PhysRevD.78.066012}
  {\bibfield  {journal} {\bibinfo  {journal} {Phys. Rev. D}\ }\textbf {\bibinfo
  {volume} {78}},\ \bibinfo {pages} {066012} (\bibinfo {year} {2008})},\
  \Eprint {https://arxiv.org/abs/0803.2876} {arXiv:0803.2876 [hep-th]}
  \BibitemShut {NoStop}%
\bibitem [{\citenamefont {Ibanez}\ and\ \citenamefont
  {Richter}(2009)}]{Ibanez:2008my}%
  \BibitemOpen
  \bibfield  {author} {\bibinfo {author} {\bibfnamefont {L.~E.}\ \bibnamefont
  {Ibanez}}\ and\ \bibinfo {author} {\bibfnamefont {R.}~\bibnamefont
  {Richter}},\ }\bibfield  {title} {\bibinfo {title} {{Stringy Instantons and
  Yukawa Couplings in MSSM-like Orientifold Models}},\ }\href
  {https://doi.org/10.1088/1126-6708/2009/03/090} {\bibfield  {journal}
  {\bibinfo  {journal} {JHEP}\ }\textbf {\bibinfo {volume} {03}},\ \bibinfo
  {pages} {090}},\ \Eprint {https://arxiv.org/abs/0811.1583} {arXiv:0811.1583
  [hep-th]} \BibitemShut {NoStop}%
\bibitem [{\citenamefont {Mayes}(2020)}]{Mayes:2019isy}%
  \BibitemOpen
  \bibfield  {author} {\bibinfo {author} {\bibfnamefont {V.~E.}\ \bibnamefont
  {Mayes}},\ }\bibfield  {title} {\bibinfo {title} {{All fermion masses and
  mixings in an intersecting D-brane world}},\ }\href
  {https://doi.org/10.1016/j.nuclphysb.2019.114848} {\bibfield  {journal}
  {\bibinfo  {journal} {Nucl. Phys. B}\ }\textbf {\bibinfo {volume} {950}},\
  \bibinfo {pages} {114848} (\bibinfo {year} {2020})},\ \Eprint
  {https://arxiv.org/abs/1902.00983} {arXiv:1902.00983 [hep-ph]} \BibitemShut
  {NoStop}%
\bibitem [{\citenamefont {Blumenhagen}\ \emph
  {et~al.}(2007{\natexlab{b}})\citenamefont {Blumenhagen}, \citenamefont
  {Cvetic},\ and\ \citenamefont {Weigand}}]{Blumenhagen:2006xt}%
  \BibitemOpen
  \bibfield  {author} {\bibinfo {author} {\bibfnamefont {R.}~\bibnamefont
  {Blumenhagen}}, \bibinfo {author} {\bibfnamefont {M.}~\bibnamefont
  {Cvetic}},\ and\ \bibinfo {author} {\bibfnamefont {T.}~\bibnamefont
  {Weigand}},\ }\bibfield  {title} {\bibinfo {title} {{Spacetime instanton
  corrections in 4D string vacua: The Seesaw mechanism for D-Brane models}},\
  }\href {https://doi.org/10.1016/j.nuclphysb.2007.02.016} {\bibfield
  {journal} {\bibinfo  {journal} {Nucl. Phys. B}\ }\textbf {\bibinfo {volume}
  {771}},\ \bibinfo {pages} {113} (\bibinfo {year} {2007}{\natexlab{b}})},\
  \Eprint {https://arxiv.org/abs/hep-th/0609191} {arXiv:hep-th/0609191}
  \BibitemShut {NoStop}%
\bibitem [{\citenamefont {Ibanez}\ and\ \citenamefont
  {Uranga}(2007)}]{Ibanez:2006da}%
  \BibitemOpen
  \bibfield  {author} {\bibinfo {author} {\bibfnamefont {L.~E.}\ \bibnamefont
  {Ibanez}}\ and\ \bibinfo {author} {\bibfnamefont {A.~M.}\ \bibnamefont
  {Uranga}},\ }\bibfield  {title} {\bibinfo {title} {{Neutrino Majorana Masses
  from String Theory Instanton Effects}},\ }\href
  {https://doi.org/10.1088/1126-6708/2007/03/052} {\bibfield  {journal}
  {\bibinfo  {journal} {JHEP}\ }\textbf {\bibinfo {volume} {03}},\ \bibinfo
  {pages} {052}},\ \Eprint {https://arxiv.org/abs/hep-th/0609213}
  {arXiv:hep-th/0609213} \BibitemShut {NoStop}%
\bibitem [{\citenamefont {Cvetic}\ \emph {et~al.}(2007)\citenamefont {Cvetic},
  \citenamefont {Richter},\ and\ \citenamefont {Weigand}}]{Cvetic:2007ku}%
  \BibitemOpen
  \bibfield  {author} {\bibinfo {author} {\bibfnamefont {M.}~\bibnamefont
  {Cvetic}}, \bibinfo {author} {\bibfnamefont {R.}~\bibnamefont {Richter}},\
  and\ \bibinfo {author} {\bibfnamefont {T.}~\bibnamefont {Weigand}},\
  }\bibfield  {title} {\bibinfo {title} {{Computation of D-brane instanton
  induced superpotential couplings: Majorana masses from string theory}},\
  }\href {https://doi.org/10.1103/PhysRevD.76.086002} {\bibfield  {journal}
  {\bibinfo  {journal} {Phys. Rev. D}\ }\textbf {\bibinfo {volume} {76}},\
  \bibinfo {pages} {086002} (\bibinfo {year} {2007})},\ \Eprint
  {https://arxiv.org/abs/hep-th/0703028} {arXiv:hep-th/0703028} \BibitemShut
  {NoStop}%
\bibitem [{\citenamefont {Cvetic}\ \emph {et~al.}(2005)\citenamefont {Cvetic},
  \citenamefont {Langacker}, \citenamefont {Li},\ and\ \citenamefont
  {Liu}}]{Cvetic:2004nk}%
  \BibitemOpen
  \bibfield  {author} {\bibinfo {author} {\bibfnamefont {M.}~\bibnamefont
  {Cvetic}}, \bibinfo {author} {\bibfnamefont {P.}~\bibnamefont {Langacker}},
  \bibinfo {author} {\bibfnamefont {T.-j.}\ \bibnamefont {Li}},\ and\ \bibinfo
  {author} {\bibfnamefont {T.}~\bibnamefont {Liu}},\ }\bibfield  {title}
  {\bibinfo {title} {{D6-brane splitting on type IIA orientifolds}},\ }\href
  {https://doi.org/10.1016/j.nuclphysb.2004.12.028} {\bibfield  {journal}
  {\bibinfo  {journal} {Nucl. Phys. B}\ }\textbf {\bibinfo {volume} {709}},\
  \bibinfo {pages} {241} (\bibinfo {year} {2005})},\ \Eprint
  {https://arxiv.org/abs/hep-th/0407178} {arXiv:hep-th/0407178} \BibitemShut
  {NoStop}%
\bibitem [{\citenamefont {Chen}\ \emph {et~al.}(2006)\citenamefont {Chen},
  \citenamefont {Li},\ and\ \citenamefont {Nanopoulos}}]{Chen:2006gd}%
  \BibitemOpen
  \bibfield  {author} {\bibinfo {author} {\bibfnamefont {C.-M.}\ \bibnamefont
  {Chen}}, \bibinfo {author} {\bibfnamefont {T.}~\bibnamefont {Li}},\ and\
  \bibinfo {author} {\bibfnamefont {D.~V.}\ \bibnamefont {Nanopoulos}},\
  }\bibfield  {title} {\bibinfo {title} {{Type IIA Pati-Salam flux vacua}},\
  }\href {https://doi.org/10.1016/j.nuclphysb.2006.01.039} {\bibfield
  {journal} {\bibinfo  {journal} {Nucl. Phys. B}\ }\textbf {\bibinfo {volume}
  {740}},\ \bibinfo {pages} {79} (\bibinfo {year} {2006})},\ \Eprint
  {https://arxiv.org/abs/hep-th/0601064} {arXiv:hep-th/0601064} \BibitemShut
  {NoStop}%
\bibitem [{\citenamefont {Chen}\ \emph {et~al.}(2008)\citenamefont {Chen},
  \citenamefont {Li}, \citenamefont {Mayes},\ and\ \citenamefont
  {Nanopoulos}}]{Chen:2007zu}%
  \BibitemOpen
  \bibfield  {author} {\bibinfo {author} {\bibfnamefont {C.-M.}\ \bibnamefont
  {Chen}}, \bibinfo {author} {\bibfnamefont {T.}~\bibnamefont {Li}}, \bibinfo
  {author} {\bibfnamefont {V.~E.}\ \bibnamefont {Mayes}},\ and\ \bibinfo
  {author} {\bibfnamefont {D.~V.}\ \bibnamefont {Nanopoulos}},\ }\bibfield
  {title} {\bibinfo {title} {{Towards realistic supersymmetric spectra and
  Yukawa textures from intersecting branes}},\ }\href
  {https://doi.org/10.1103/PhysRevD.77.125023} {\bibfield  {journal} {\bibinfo
  {journal} {Phys. Rev. D}\ }\textbf {\bibinfo {volume} {77}},\ \bibinfo
  {pages} {125023} (\bibinfo {year} {2008})},\ \Eprint
  {https://arxiv.org/abs/0711.0396} {arXiv:0711.0396 [hep-ph]} \BibitemShut
  {NoStop}%
\bibitem [{\citenamefont {Navas}\ \emph {et~al.}(2024)\citenamefont {Navas}
  \emph {et~al.}}]{ParticleDataGroup:2024cfk}%
  \BibitemOpen
  \bibfield  {author} {\bibinfo {author} {\bibfnamefont {S.}~\bibnamefont
  {Navas}} \emph {et~al.} (\bibinfo {collaboration} {Particle Data Group}),\
  }\bibfield  {title} {\bibinfo {title} {{Review of particle physics}},\ }\href
  {https://doi.org/10.1103/PhysRevD.110.030001} {\bibfield  {journal} {\bibinfo
   {journal} {Phys. Rev. D}\ }\textbf {\bibinfo {volume} {110}},\ \bibinfo
  {pages} {030001} (\bibinfo {year} {2024})}\BibitemShut {NoStop}%
\bibitem [{\citenamefont {Bona}\ \emph {et~al.}(2023)\citenamefont {Bona} \emph
  {et~al.}}]{UTfit:2022hsi}%
  \BibitemOpen
  \bibfield  {author} {\bibinfo {author} {\bibfnamefont {M.}~\bibnamefont
  {Bona}} \emph {et~al.} (\bibinfo {collaboration} {UTfit}),\ }\bibfield
  {title} {\bibinfo {title} {{New UTfit Analysis of the Unitarity Triangle in
  the Cabibbo-Kobayashi-Maskawa scheme}},\ }\href
  {https://doi.org/10.1007/s12210-023-01137-5} {\bibfield  {journal} {\bibinfo
  {journal} {Rend. Lincei Sci. Fis. Nat.}\ }\textbf {\bibinfo {volume} {34}},\
  \bibinfo {pages} {37} (\bibinfo {year} {2023})},\ \Eprint
  {https://arxiv.org/abs/2212.03894} {arXiv:2212.03894 [hep-ph]} \BibitemShut
  {NoStop}%
\bibitem [{\citenamefont {Montero}\ \emph {et~al.}(2023)\citenamefont
  {Montero}, \citenamefont {Vafa},\ and\ \citenamefont
  {Valenzuela}}]{Montero:2022prj}%
  \BibitemOpen
  \bibfield  {author} {\bibinfo {author} {\bibfnamefont {M.}~\bibnamefont
  {Montero}}, \bibinfo {author} {\bibfnamefont {C.}~\bibnamefont {Vafa}},\ and\
  \bibinfo {author} {\bibfnamefont {I.}~\bibnamefont {Valenzuela}},\ }\bibfield
   {title} {\bibinfo {title} {{The dark dimension and the Swampland}},\ }\href
  {https://doi.org/10.1007/JHEP02(2023)022} {\bibfield  {journal} {\bibinfo
  {journal} {JHEP}\ }\textbf {\bibinfo {volume} {02}},\ \bibinfo {pages}
  {022}},\ \Eprint {https://arxiv.org/abs/2205.12293} {arXiv:2205.12293
  [hep-th]} \BibitemShut {NoStop}%
\bibitem [{\citenamefont {Lee}\ \emph {et~al.}(2020)\citenamefont {Lee},
  \citenamefont {Adelberger}, \citenamefont {Cook}, \citenamefont {Fleischer},\
  and\ \citenamefont {Heckel}}]{Lee:2020zjt}%
  \BibitemOpen
  \bibfield  {author} {\bibinfo {author} {\bibfnamefont {J.~G.}\ \bibnamefont
  {Lee}}, \bibinfo {author} {\bibfnamefont {E.~G.}\ \bibnamefont {Adelberger}},
  \bibinfo {author} {\bibfnamefont {T.~S.}\ \bibnamefont {Cook}}, \bibinfo
  {author} {\bibfnamefont {S.~M.}\ \bibnamefont {Fleischer}},\ and\ \bibinfo
  {author} {\bibfnamefont {B.~R.}\ \bibnamefont {Heckel}},\ }\bibfield  {title}
  {\bibinfo {title} {{New Test of the Gravitational $1/r^2$ Law at Separations
  down to 52 $\mu$m}},\ }\href {https://doi.org/10.1103/PhysRevLett.124.101101}
  {\bibfield  {journal} {\bibinfo  {journal} {Phys. Rev. Lett.}\ }\textbf
  {\bibinfo {volume} {124}},\ \bibinfo {pages} {101101} (\bibinfo {year}
  {2020})},\ \Eprint {https://arxiv.org/abs/2002.11761} {arXiv:2002.11761
  [hep-ex]} \BibitemShut {NoStop}%
\end{thebibliography}
 
\end{document}